\documentstyle[twoside,psfig]{article}

\catcode`\@=11 \long\def\@makefntext#1{ \protect\noindent \hbox to
3.2pt {\hskip-.9pt
$^{{\eightrm\@thefnmark}}$\hfil}#1\hfill}       %CAN BE USED

\def\@makefnmark{\hbox to 0pt{$^{\@thefnmark}$\hss}}    %ORIGINAL

\def\ps@myheadings{\let\@mkboth\@gobbletwo
\def\@oddhead{\hbox{}
\rightmark\hfil\eightrm\thepage}
\def\@oddfoot{}\def\@evenhead{\eightrm\thepage\hfil
\leftmark\hbox{}}\def\@evenfoot{}
\def\sectionmark##1{}\def\subsectionmark##1{}}

\oddsidemargin=\evensidemargin 
\newcounter{appendixc}
\newcounter{subappendixc}[appendixc]
\newcounter{subsubappendixc}[subappendixc]
\renewcommand{\thesubappendixc}{\Alph{appendixc}.\arabic{subappendixc}}
\renewcommand{\thesubsubappendixc}
    {\Alph{appendixc}.\arabic{subappendixc}.\arabic{subsubappendixc}}

\renewcommand{\appendix}[1] {\vspace{12pt}
        \refstepcounter{appendixc}
        \setcounter{figure}{0}
        \setcounter{table}{0}
        \setcounter{lemma}{0}
        \setcounter{theorem}{0}
        \setcounter{corollary}{0}
        \setcounter{definition}{0}
        \setcounter{equation}{0}
        \renewcommand{\thefigure}{\Alph{appendixc}.\arabic{figure}}
        \renewcommand{\thetable}{\Alph{appendixc}.\arabic{table}}
        \renewcommand{\theappendixc}{\Alph{appendixc}}
        \renewcommand{\thelemma}{\Alph{appendixc}.\arabic{lemma}}
        \renewcommand{\thetheorem}{\Alph{appendixc}.\arabic{theorem}}
        \renewcommand{\thedefinition}{\Alph{appendixc}.\arabic{definition}}
        \renewcommand{\thecorollary}{\Alph{appendixc}.\arabic{corollary}}
        \noindent{\tenbf Appendix \theappendixc #1}\par\vspace{5pt}}
\newcommand{\subappendix}[1] {\vspace{12pt}
        \refstepcounter{subappendixc}
        \noindent{\bf Appendix \thesubappendixc. {\kern1pt \bfit #1}}
    \par\vspace{5pt}}
\newcommand{\subsubappendix}[1] {\vspace{12pt}
        \refstepcounter{subsubappendixc}
        \noindent{\rm Appendix \thesubsubappendixc. {\kern1pt \tenit #1}}
    \par\vspace{5pt}}

\newcommand{\textlineskip}{\baselineskip=13pt}
\newcommand{\smalllineskip}{\baselineskip=10pt}

\renewenvironment{thebibliography}[1]
        {\frenchspacing
     \ninerm\baselineskip=11pt
         \begin{list}{\arabic{enumi}.}
        {\usecounter{enumi}\setlength{\parsep}{0pt}
     \setlength{\leftmargin 12.7pt}{\rightmargin 0pt} %FOR 1--9 ITEMS
         \setlength{\itemsep}{0pt} \settowidth
    {\labelwidth}{#1.}\sloppy}}{\end{list}}

\newcounter{itemlistc}
\newcounter{romanlistc}
\newcounter{alphlistc}
\newcounter{arabiclistc}

\newcommand{\fcaption}[1]{
        \refstepcounter{figure}
        \setbox\@tempboxa = \hbox{\footnotesize Fig.~\thefigure. #1}
        \ifdim \wd\@tempboxa > 5in
           {\begin{center}
        \parbox{5in}{\footnotesize\smalllineskip Fig.~\thefigure. #1}
            \end{center}}
        \else
             {\begin{center}
             {\footnotesize Fig.~\thefigure. #1}
              \end{center}}
        \fi}

\newcommand{\tcaption}[1]{
        \refstepcounter{table}
        \setbox\@tempboxa = \hbox{\footnotesize Table~\thetable. #1}
        \ifdim \wd\@tempboxa > 5in
           {\begin{center}
        \parbox{5in}{\footnotesize\smalllineskip Table~\thetable. #1}
            \end{center}}
        \else
             {\begin{center}
             {\footnotesize Table~\thetable. #1}
              \end{center}}
        \fi}

\def\@citex[#1]#2{\if@filesw\immediate\write\@auxout
    {\string\citation{#2}}\fi
\def\@citea{}\@cite{\@for\@citeb:=#2\do
    {\@citea\def\@citea{,}\@ifundefined
    {b@\@citeb}{{\bf ?}\@warning
    {Citation `\@citeb' on page \thepage \space undefined}}
    {\csname b@\@citeb\endcsname}}}{#1}}

\newif\if@cghi
\def\cite{\@cghitrue\@ifnextchar [{\@tempswatrue
    \@citex}{\@tempswafalse\@citex[]}}
\def\citelow{\@cghifalse\@ifnextchar [{\@tempswatrue
    \@citex}{\@tempswafalse\@citex[]}}
\def\@cite#1#2{{$\null^{#1}$\if@tempswa\typeout
    {IJCGA warning: optional citation argument
    ignored: `#2'} \fi}}

\def\pmb#1{\setbox0=\hbox{#1}
    \kern-.025em\copy0\kern-\wd0
    \kern.05em\copy0\kern-\wd0
    \kern-.025em\raise.0433em\box0}

\def\fnt#1#2{\footnotetext{\kern-.3em
    {$^{\mbox{\scriptsize #1}}$}{#2}}}

\def\ps@myheadings{%
    \let\@oddfoot\@empty\let\@evenfoot\@empty
    \def\@evenhead{\slshape\leftmark\hfil}%       %EVEN PAGE
    \def\@oddhead{\hfil{\slshape\rightmark}}%     %ODD PAGE
    \let\@mkboth\@gobbletwo
    \let\sectionmark\@gobble
    \let\subsectionmark\@gobble
    }
 \font\tenit=cmti10 \font\tenbf=cmbx10
\font\bfit=cmbxti10 at 10pt \font\ninerm=cmr9 
 \font\eightrm=cmr8

\def\qed{\hbox{${\vcenter{\vbox{            %HOLLOW SQUARE
   \hrule height 0.4pt\hbox{\vrule width 0.4pt height 6pt
   \kern5pt\vrule width 0.4pt}\hrule height 0.4pt}}}$}}

\def\bsc{{\sc a\kern-6.4pt\sc a\kern-6.4pt\sc a}}   %LATEX LOGO
\def\bflatex{\bf L\kern-.30em\raise.3ex\hbox{\bsc}\kern-.14em
T\kern-.1667em\lower.7ex\hbox{E}\kern-.125em X}

\begin{document}
\setlength{\textheight}{7.7truein}  %for 2nd page onwards

\thispagestyle{empty}

\normalsize\textlineskip

\setcounter{page}{1}

%\copyrightheading{}         %{Vol. 0, No. 0 (1993) 000--000}

\vspace*{0.88truein}

%\fpage{1}
\centerline{\bf SUMMARY OF THE LATIN AMERICAN WORKSHOP\\}
\vspace*{0.035truein} \centerline{\bf ON FUNDAMENTAL
INTERACTIONS\footnote{Partially supported by CONICET and ANPCyT
-Argentina.}} \vspace*{0.37truein}
\vspace*{15pt}          %when needed
\centerline{\footnotesize C.A. GARC\'{I}A CANAL}
\baselineskip=12pt \centerline{\footnotesize\it Laboratorio de
F\'{\i}sica Te\'{o}rica, Departamento de F\'{\i}sica, Universidad
Nacional de La Plata} \baselineskip=10pt
\centerline{\footnotesize\it C.C. 67 - 1900, La Plata, Argentina}
\centerline{\footnotesize\it E-mail: garcia@fisica.unlp.edu.ar}
\vspace*{15pt}

 \vspace*{0.225truein}
\vspace*{15pt}

\begin{abstract}

 Summary of the Latin American Workshop on the
Fundamental Interactions held at the Physics Department of the
Universidad de Buenos Aires from 26 to 30 July 2004.
\end{abstract}

\pagebreak

\section*{Introduction}
Simultaneously with the Sixth J.J. Giambiagi Winter School of
Physics \cite{School} organized by the Physics Department of the
Buenos Aires University, a new edition of the Latin American
Workshop on the Fundamental Interactions took place. The workshop
was organized in association with the Latin American Network on
Phenomenology of the Fundamental Interactions. As in previous
editions of the workshop, it aimed to tighten the already existing
links between the High Energy Physics community of the region. The
workshop offered the possibility of presenting recent results and
research in progress to many members of the community, both young
scientists and more experienced researchers.

There were over 30 contributions presented by participants coming
from 19 institutions from 9 different countries. These figures
speak by themselves about the impact of the meeting.

At a first sight one can divide the mentioned contributions among
9 different topics, namely: Quantum Chromodynamics with 6
contributions, Beyond the Standard Model, Hadrons and Cosmic Rays
Physics with 4, Neutrino Physics, CP Violation and Chiral Models
with 2, Astroparticles with 1 contribution and Strings and Field
Theory with 7. This distribution of topics provides, in some
sense, a panorama of the main interest of the research activities
in progress in the region.

Before entering in a more detailed analysis of the work presented
during the workshop, we would like to stress not only the high
level of the presentations but the actual interest of the subjects
covered. No doubt, all the topics treated are of real present
interest to the international community. The results discussed
have been recently presented in renowned international
publications or they will be published in the near future. Another
point that has to be emphatically remarked is the young condition
of the almost two thirds of the active participants in the
workshop. All these details allow one to ensure that a promising
future can be expected for the phenomenology of the fundamental
interactions in the region.

\section*{Summary Report}

\subsection*{Quantum Chromodynamics}

\subsubsection*{The photon structure function: new parameterizations
for the parton content of the photon  \cite{Cornet}}

New NLO parameterizations for the parton content of the real
photon were presented. These parameterizations were obtained
performing fits to all the available data for the Photon Structure
Function with $Q^2>1\, GeV^2$. Special attention was given to the
way of dealing with the heavy quark contributions and various
approaches were compared.

\subsubsection*{Single spin asymmetries \cite{Schmidt}}

Among the very interesting developments of spin physics in last
years, the azimuthal asymmetry found originally by the Hermes
collaboration in pion electroproduction is particularly
outstanding. The importance of these measurements was stressed on
the basis that they represent a new class of measurements of spin
processes, and act as a filter for exotic parton distributions
such as transversity.

\subsubsection*{Dilepton $p_T$ distribution through the Color Glass
Condensate \cite{Gay}}

The dilepton production in proton-nucleus collisions at the
forward region was investigated using  the  Color Glass
Condensate, a saturated dense system of partons, particularly
gluons. In the high energy kinematical limit the saturation of the
gluon density is required by unitarity. It was found that the
lepton pair is a most suitable observable to analyze this
saturated regime. It was shown that saturation effects are large
in the transverse momentum distribution ($p_T$) of the dileptons,
particularly in the low $p_T$ region. The calculations were
performed for the energies of interest at RHIC ($\sqrt{s}=350 \,
GeV$) and LHC ($\sqrt{s}=8800 \, GeV$).

\subsubsection*{Consistency of NLO global analysis \cite{Navarro}}

A detailed study of consistency between different sets of
polarized deep inelastic scattering data and theory was presented.
The analysis followed the criteria proposed by Collins and
Pumplin. It was also discussed the implementation of the double
Mellin transform into the global fit, in order to circumvent the
extremely time-consuming convolution integrals involved in the
standard calculation.

\subsubsection*{High density Quantum Chromodynamics through hard probes
\cite{Betemps}}

Investigations on the signature of the Color Glass Condensate,
saturated dense system of partons (gluons), were presented. The
signature of these non-linear effects in QCD were analyzed by
considering appropriate observables carrying suitable information.
Specific distributions of the observables of the saturation
effects, in particular the transverse momentum of the dilepton
produced in the heavy ion colliders, were analyzed.

\subsubsection*{Meson production in ultrarelativistic heavy ion collisions
\cite{Mackedanz}}

A critical analysis of the signature of deconfined strongly
interacting matter, the Quark-Gluon Plasma (QGP), was presented.
As the QGP can only be detected indirectly, one of the possible
signatures is the quarkonium suppresssion, in particular the J/Psi
suppresion. It was discussed that mesons are not only affected by
the formation of the deconfined phase because other possible
contributions to the suppression are present. In the initial
state, effects as shadowing, partonic energy loss before the hard
interaction and the parton recombination, due to the high partonic
density, were taken into account. In the final state, rescattering
effects and mechanisms associated with the high density produced
in the collisions were considered. Hadron production at RHIC was
analyzed in order to determine the magnitude  of the initial state
effects present in the corresponding kinematical regime.

\subsection*{Beyond the Standard Model}

\subsubsection*{Double charged bileptons in $3-3-1$ models
\cite{Ramirez}}

The production of bileptons within the model with $SU(3) \otimes
SU(3) \otimes U(1)$ gauge symmetry was studied. This model
predicts single and double charged bileptons and exotic quarks
carrying charges of $-4/3$ and $5/3$ units of the positron charge.
The analysis of the phenomenological consequences of the model was
done in connection with $e^+ \,e^-$ and $p\,p$ collisions at
future linear colliders and LHC.

\subsubsection*{Study of new gauge boson properties at ATLAS \cite{Nepomuceno}}

A possible signal and properties of a new neutral gauge boson
$Z'$, in a mass range $800 - 2000 \, GeV$, were investigated in
the channel proton + proton -> muon + muon + photon, at LHC for CM
energy of $14 \, TeV$ and a luminosity of $100/fb$. CompHep,
PYTHIA and detector simulation packages were used for the
calculation. The total cross-sections, mass distributions,
transversal momentum and asymmetry were analyzed in order to
identify and distinguish the two minimal left-right mirror and
left-right symmetric models. The preliminary simulation results
have shown that the $Z'$ could be observed in the above channel
and its properties determined by fitting some muon distributions.

\subsubsection*{Finite temperature behavior of an extended model
with a triplet Higgs boson \cite{Moreno}}

The finite temperature effective potential of an extension of the
Standard Model with an extra triplet Higgs boson was computed.
Constraints on the parameters of the model consistent with
precision electroweak measurements and cosmology were determined.

\subsubsection*{Spontaneous parity breaking and fermion masses
\cite{Simoes}}

A $SU(2)_{L} \otimes SU(2)_{R} \otimes  U(1)$ left-right symmetric
model for elementary particles was presented and its connection
with the fermion mass spectrum was considered. The model includes
new mirror fermions and a minimal set of Higgs particles and can
accommodate a consistent pattern for charged and neutral fermion
masses as well as neutrino oscillations. The connection between
the left and right sectors is done through the neutral vector
gauge boson $Z$ and a new heavier boson $Z'$.

\subsection*{Hadrons}

\subsubsection*{A new signature for glueballs \cite{Lopes}}

An effective glueball-glueball interaction Hamiltonian in the
context of a constituent gluon model, involving explicit gluon
degrees of freedom was derived by means of a mapping technique
(the Fock-Tani formalism). The calculation was focused on the
interacting $J^{PC}=0^{++}$ state. The $0^{++}$ resonance, being
an isospin zero state can be either represented as a $s\bar{s}$
bound state or a glueball with the same quantum numbers. The
different values of the corresponding cross-section were proposed
as a mean for distinguishing between both descriptions.

\subsubsection*{The tower structure of L=1 excited baryons and
their strong decays \cite{Schat}}

In the large $N_c$ limit, the mass spectrum of the $L=1$ orbitally
excited baryons $N*$, $\Delta^*$ is shown to have a very simple
structure, with states degenerate in pairs of spins $J=(1/2, 3/2),
(3/2,5/2)$,corresponding to irreducible representations (towers)
of the contracted $SU(4)_c$ symmetry group. The mixing angles were
completely determined in this limit. The $1/N_c$ corrections were
computed, pointing out a four-fold ambiguity in the correspondence
of the observed baryons with the large $N_c$ states. The leading
order predictions for the strong decays were also discussed.

\subsubsection*{Annihilation amplitudes and factorization in $B^{+/-}$ to $\phi
{K^*}^{+/-}$\cite{Szyn}}

The decay $B^{+/-}$ to $\phi {K^*}^{+/-}$, followed by the decay
of the outgoing vector mesons into two pseudoscalars was studied.
The analysis of angular distributions of the decay products was
shown to provide useful information about the annihilation
contributions and possible tests of factorization.

\subsubsection*{Pion nucleon scattering and pion photoproduction. Dynamical
models \cite{Mariano}}

The phenomenological treatment of unstable particles in reactions
generated by hadron and electromagnetic probes on the nucleon was
presented. As a consequence, a consistent model that respects
gauge invariance, Lorentz invariance and invariance under contact
transformations in the spin-3/2 fields was obtained.

\subsection*{Cosmic Rays Physics}

\subsubsection*{Double extensive air shower induced by ultra-high
energy cosmic tau-neutrino \cite{Adrega}}

 The role of ultra-high energy cosmic neutrinos in explaining
the origin of cosmic rays with energies beyond the GZK limit of
few times $10^19\, eV$ was considered. As neutrinos hardly
interact with cosmic microwave background or intergalactic
magnetic fields, they keep their original energy and direction of
propagation. When neutrino flavor oscillations are taken into
consideration the flavor proportion is
$\nu_e:\nu_\mu:\nu_\tau~1:1:1$ instead of
$\nu_e:\nu_\mu:\nu_\tau=1:2:0$ as expected a priori. On this
basis, the possibility of detecting ultra-high energy cosmic
tau-neutrinos was analyzed by means of a process involving a
double extensive air shower, the so-called Double-Bang Phenomenon.
In this process a primary tau-neutrino interacts  with an
atmospheric quark creating a first hadronic shower and a
tau-lepton,  which subsequently decays creating a second cascade.
As the number of these events strongly depends on the flux of
tau-neutrinos arriving at the Earth's atmosphere, they can be used
to test theoretical models for the production of ultra-high energy
tau-neutrinos. The potential of the fluorescence detector of the
Pierre Auger Observatory to observe Double-Bang events was
studied.

\subsubsection*{Revision of the neutrinos oscillation probability in
the supernovae \cite{Aliaga}}

The analytical expression of the oscillations probability in the
supernovae and the use of a simple analytical prescription was
reviewed. A comparison with the numerical results which are
obtained from the solution of the evolution equation was also
performed.

\subsubsection*{Absolute calibration of Auger fluorescence
detector\cite{Tamashiro}}

The description of the calibrated light device called Drum,
designed and used at the Fluorescence Detector of the Pierre Auger
Observatory was presented. It is being used for the absolute
end-to-end calibration of each of its telescopes. The Drum is a
2.5m diameter device that simulates a far away point source of UV
light pulses (shower like) that can be mounted on the aperture of
any Fluorescence Detector telescopes of the Observatory. The
instrument can be taken as a black-box because when injecting a
known input flux and measuring its response under normal operation
conditions, the calibration constant can be calculated for each of
its pixels (photo-multipliers).

\subsubsection*{Calibration of the Pierre Auger surface detectors
\cite{Bonifazi}}

The remoteness and difficulty to access each detector of the
ground array of the Pierre Auger Observatory that consists of 1600
Water Cerenkov Detectors deployed over 3000 km2 at Pampa Amarilla,
near Malargüe, in Mendoza, Argentina, imposed an automatic remote
calibration procedure that was designed. It was decided that the
main parameter to be determined is the average charge deposited by
central muons crossing the water volume vertically.

\subsection*{Neutrino Physics}

\subsubsection*{Neutrino factories and the degeneracies in the
measurement of neutrino oscillation parameters \cite{Jones}}

 The future neutrino factories and their characteristics for
measuring neutrino oscillation parameters contained in the
Pontecorvo-Maki-Nakagawa-Sakata mixing matrix were reviewed. The
degeneracies in neutrino oscillation parameters was analyzed,
making a quantitative analysis to find out how to solve these
degeneracies in neutrino factories. To this end, the so called
"Golden" and "Silver" channels, combining the effect of measuring
particles and antiparticles in the same experiment were
considered.

\subsubsection*{Neutrino absorption tomography of the Earth
\cite{Reynoso}}

The passage of Ultra High Energy neutrinos through the Earth was
studied in order to perform an absorption tomography of its inner
structure. The Earth's density was reconstructed using the 2-d
Radon transform considering neutral current regeneration through
the complete transport equation and the results were compared with
the ones obtained with the effective cross section. It was found
that the transport equation approach is more accurate than the
effective cross section approach especially for flat initial
spectra, and that the recovered density presents a percentage
uncertainty less than two times the uncertainty in the flux.

\subsection*{CP Violation}

\subsubsection*{Correlated B-meson decays into CP-eigenstates \cite{Alvarez}}

It was shown that in the B-meson correlated decay  experiment the
decays to $J/\Psi K_{SL}$ can be used to place CP-tags.  This
theoretical tool allows the correlated decays analysis of
Flavor-Flavor and Flavor-CP to be completed with the decays
CP-Flavor  and CP-CP.  These results show a way to measure
Standard Model  parameters, as for instance sin (2 beta), and
through  CP $\Delta t$ asymmetries, an observable can be
introduced to obtain $\Delta \Gamma$ linearly.

\subsubsection*{CP and T Trajectory diagrams for a unified
graphical representation in context of neutrino oscillations
\cite{Ribeiro}}

The CP trajectory diagrams, introduced to demonstrate the
difference between the intrinsic CP violating effects and those
induced by matter for neutrino oscillation were analyzed.

\subsection*{Chiral Models}

\subsubsection*{Two flavor superconductivity in nonlocal chiral quark
models \cite{Scoccola}}

A relativistic quark model with non-local quark-antiquark and
quark-quark interactions at finite temperature and density was
studied. The structure of the corresponding phase diagram was
analyzed and the competition between the chiral and two
superconducting phases discussed.

\subsubsection*{Properties of light pseudoscalar mesons in a nonlocal
chiral quark model \cite{Gomez}}

The predictions for masses and decay constants of light
pseudoscalar mesons in a nonlocal $SU(3)$ chiral quark model were
presented. The results found are in reasonable agreement with the
corresponding empirical values, in particular in the case of the
ratio $f_K/f_pi$, the decay $pi^0 \rightarrow \gamma \gamma$ and
the observed phenomenology in the $\eta-\eta'$ sector. The
$\eta-\eta'$ mixing was described by introducing two mixing angles
which turn out to be significantly different from each other.

\subsection*{Astroparticles}

\subsubsection*{Pulsar velocities and active-sterile neutrino
oscillations\cite{Dolivo}}

Active-sterile neutrino transformations in a magnetized
protoneutron star were described on the basis of a spherical
resonance layer for neutrinos moving in different directions. The
asymmetry in the momentum taken away by neutrinos was expressed in
terms of the thickness of the resonance shell, which depends on
the intensity of the magnetic field.  The required pulsar kicks
could be obtained with the magnetic fields expected in the
interior of a protoneutron star, for sterile neutrino masses of
the order of $keV$ and small mixing angles.

\subsection*{Strings and Field Theory}

\subsubsection*{Composite states in string theory: an holographic
view\cite{Giribet}}

The contribution of composite (two-string) states in the
intermediate channels of the four-point scattering processes in
Anti-de Sitter space were identified. By means of the study of
Maldacena AdS/CFT holographic description of the processes in AdS
spacetime the factorization properties of the scattering
amplitudes were analyzed.

\subsubsection*{The CPT group of the Dirac field \cite{Soco}}

By using the standard representation of the Dirac equation it was
shown that, up to signs, there exist only two sets of consistent
solutions for the matrices of charge conjugation (C), parity (P),
and time reversal (T). In both cases, $P^2=-1$, and then two
successive applications of the parity transformation to spin 1/2
fields necessarily amounts to a 2 rotation. Each of these sets
generates a non abelian group of sixteen elements, G1 and G2, non
isomorphic subgroups of the Dirac algebra. G1 is isomorphic to
$D8xZ2$, where $D8$ is the dihedral group of  eight elements (the
symmetries of the square); while G2 is isomorphic to a certain
semidirect product of $D8$ and $Z2$. On the other hand, the
corresponding quantum operators for C, P, and T generate a unique
group G, called the CPT group of the Dirac field, isomorphic to
$QxZ2$, where $Q$ is the quaternion group.  The matrix groups were
given in the Weyl and Majorana representations.

\subsubsection*{Classical limit of the scattering of Dirac particles by magnetic
fields\cite{Murguia}}

The relativistic quantum scattering of charged fermions with a
solenoidal magnetic field was studied. The analysis was done in
the frame of the perturbation theory with free particle asymptotic
states. The classical limit of the resulting cross section was
analyzed, finding a consistency with the well known Aharonov-Bhom
result, that corresponds to the zero radii limit of the solenoid.
The large solenoid radii limit, situation that corresponds to the
interaction with a constant and uniform magnetic field was also
studied.

\subsubsection*{Clifford algebras and spinors \cite{Cervantes}}

 The general classification of real and complex Clifford algebras
of finite dimension was presented. The naturality of this
classification and some applications to the spinor theory was
discussed.

\subsubsection*{The Thirring-Wess model revisited: a functional integral
approach \cite{Rodrigues}}

In order to obtain the functional integral bosonization of the
Thirring-Wess model with an arbitrary regularization parameter,
the Wess-Zumino-Witten theory was considered.By decomposing the
Bose field algebra into gauge-invariant and gauge-noninvariant
field subalgebras, the local decoupled quantum action was
obtained. The isomorphism between the QED2 (QCD2) with broken
gauge symmetry by a regularization prescription and the Abelian
(non-Abelian) Thirring-Wess model with a fixed bare mass for the
meson field was established.

\subsubsection*{Matrix models and Quantum Hall effects \cite{Ivan}}

The matrix model proposed by Polychronakos that is supposed to
realize Susskind conjecture and that establishes a one-to-one
correspondence between non-commutative Chern-Simons theories and
Laughlin's states was reviewed. Some issues related to gauge
invariance and the physical properties of observable charges and
currents were questioned.

\subsubsection*{Consistent discretization of constrained theories \cite{Ponce}}

Consistent discretization of the Gambini-Pullin canonical
formalism to treat constrained theories in Lorentz signature
space-times was presented. In this way, constraints and evolution
equations can be solved simultaneously up to small deviations
proportional to the step of the discretization.

\section*{Final Remarks}

The First Latin American Workshop on Phenomenology of the
Fundamental Interactions was held at Oaxtepec - M\'{e}xico in
December 1990. At that time, Roberto Peccei from UCLA was invited
to present the summary of the meeting. In the Proceedings one
reads as his final sentence: "Phenomenology is alive in Latin
America". We are sure that accordingly, we can end our summary of
the Workshop in Buenos Aires by stating: Phenomenology of the
Fundamental Interactions has certainly grown in Latin America. It
has a healthy life and deserves a long life.

\subsection*{Acknowledgment}
The workshop was possible due to the generous help of Universidad
de Buenos Aires (UBA), Centro Latinoamericano de F\'{\i }sica
(CLAF), International Centre for Theoretical Physics (ICTP),
Fundaci\'{o}n Sauberan, Fundaci\'{o}n Antorchas, Fundaci\'{o}n
Bunge y Born and Academia Nacional de Ciencias Exactas, F\'{\i
}sicas y Naturales de Argentina.

The essential and exhausting work of Daniel de Florian and Rodolfo
Sassot and the very important contribution of Gabriela Navarro in
organizing the Workshop is warmly acknowledged by all the
participants.

\end{document}